\documentclass[%
 reprint,
 amsmath,amssymb,
 aps,
 reprint,
 superscriptaddress,
pra,
]{revtex4-2}

\usepackage{graphicx}%
\usepackage{dcolumn}%
\usepackage{bm}%
\usepackage{siunitx}
\usepackage{amsmath, mathtools, physics}
\usepackage{array}
\usepackage{complexity}
\usepackage{tikz}
\usepackage{quantikz}
\usepackage{tikz-3dplot} 
\usepackage{booktabs}
\usepackage{soul}
\usetikzlibrary{%
    decorations.pathreplacing,%
    decorations.pathmorphing%
}
\usetikzlibrary{calc}
\usetikzlibrary{math}
\usepackage{cleveref}

\usepackage{xcolor}

\DeclareMathOperator*{\argmax}{argmax}

\definecolor{niceblue}{HTML}{a5c8e1}
\definecolor{niceorange}{HTML}{ffcb9e}

\newlength{\Ubraceignoreleft}
\newlength{\Ubraceignoreright}

\NewDocumentCommand{\Ubrace}{ommo}{%
  {%
   \IfValueT{#1}{\settowidth{\Ubraceignoreleft}{$#1$}}%
   \IfValueT{#4}{\settowidth{\Ubraceignoreright}{$#4$}}%
   \hspace*{\Ubraceignoreleft}
   \underbrace{%
     \hspace*{-\Ubraceignoreleft}%
     #2%
     \hspace*{-\Ubraceignoreright}%
   }_{#3}%
   \hspace*{\Ubraceignoreright}%
  }%
}

\graphicspath{{arxiv-figs/}}

\begin{document}

\title{Resilience of the surface code to error bursts}%

\author{Shi Jie Samuel Tan}
\affiliation{Joint Center for Quantum Information and Computer Science, University of Maryland, College Park, MD, USA.}
\affiliation{Department of Computer Science, University of Maryland, College Park, MD, USA.}
\affiliation{Haverford College, Haverford, PA 19041}
\author{Christopher A. Pattison}
\affiliation{Institute for Quantum Information and Matter, California Institute of Technology, Pasadena, CA, USA.}
\author{Matt McEwen}
\affiliation{Google Quantum AI, Santa Barbara, CA 93111, USA}
\author{John Preskill}
\affiliation{Institute for Quantum Information and Matter, California Institute of Technology, Pasadena, CA, USA.}
\affiliation{AWS Center for Quantum Computing, Pasadena, CA, USA.}

\date{\today}

\begin{abstract}
Quantum error correction works effectively only if the error rate of gate operations is sufficiently low.
However, some rare physical mechanisms can cause a temporary increase in the error rate that affects many qubits; examples include ionizing radiation in superconducting hardware and large deviations in the global control of atomic systems.
We refer to such rare transient spikes in the gate error rate as error bursts.
In this work, we investigate the resilience of the surface code to generic error bursts.
We assume that, after appropriate mitigation strategies,
the spike in the error rate lasts for only a single syndrome extraction cycle; we also assume that the enhanced error rate is uniform across the code block. Under these assumptions, and for a circuit-level depolarizing noise model, we perform Monte Carlo simulations to determine the regime in burst error rate and background error rate for which the memory time becomes arbitrarily long as the code block size grows. Our results indicate that suitable hardware mitigation methods combined with standard decoding methods may suffice to protect against transient error bursts in the surface code. 
\end{abstract}

\maketitle

\section{Introduction}%

Because quantum hardware is intrinsically prone to error, a large-scale quantum computer will need to be robust against noise. This can be accomplished by executing an encoded version of a quantum circuit, where every qubit is replaced by a logical qubit protected by a quantum error-correcting code (QECC).

If the error rate does not exceed a critical value, the \emph{accuracy threshold}, the output of an ideal quantum circuit can be sampled with arbitrary accuracy by a noisy fault-tolerant circuit if the code block is sufficiently large.
The most studied candidate QECC for achieving fault-tolerance, the surface code \cite{bravyi1998quantum,dennis2002topological}, has a high threshold and may be implemented using nearest-neighbor connectivity on a two-dimensional lattice.
These features make the surface code well suited for implementation in a solid state platform such as a superconducting circuit with transmon qubits \cite{koch2007charge,blais2021circuit}.

Recent experimental studies of superconducting quantum processors \cite{google2023suppressing,cardani2023disentangling,mcewen2022resolving} suggest that ionizing radiation may pose serious limitations on the size of quantum circuits that can be executed fault tolerantly using these devices. Such ionizing radiation, for example a cosmic ray muon, can create many quasiparticles, significantly reducing the qubit T1 time, and hence greatly enhancing the error rate in a large spatial region of the device. Because many qubits in a single code block may be affected, this event can cause an uncorrectable logical error. To avoid this catastrophe, one might place the quantum computer deep underground to suppress the muon flux, or use a hierarchical error correction scheme which protects against events that damage many qubits in a subblock of a larger code block \cite{xu2022distributed,pattison2023hierarchical}. 
Here we consider a third option: We envision that hardware mitigation methods such as quasiparticle traps \cite{martinis2021saving} limit the temporal extent of the ionization event, and investigate whether a transient spike in the error rate can be tolerated by the surface code. 

Such error bursts may occur in other platforms besides superconducting circuits, and for other reasons besides ionizing radiation. For example, brief episodes of elevated gate error rates might arise from rare large deviations from average behavior in global control systems. Here too suitable mitigation strategies can alleviate such global control noise. Our results provide guidance concerning how much mitigation is required to execute long fault-tolerant quantum computations with high success probability.

\subsection{Sustainable threshold and burst threshold}
To achieve fault tolerance using surface codes, logical operations are interleaved with so-called \emph{recovery operations}.
In each recovery operation, the stabilizer generators \cite{Gottesman_1997} that define the code subspace are projectively measured using a \emph{syndrome extraction circuit}.
Although syndrome extraction is noisy and therefore might introduce additional errors, error correction succeeds with high probability if the error rate is below threshold and the code block is sufficiently large.
We refer to a single execution of a syndrome extraction circuit producing one measurement outcome for each stabilizer generator as a \emph{round} of syndrome extraction.
For the surface code, in the presence of measurement errors, it is necessary to repeat the syndrome extraction circuit a number of times roughly equal to the code distance $d$ \cite{dennis2002topological}.
We refer to this $O(d)$-times repetition of the syndrome extraction circuit as one \emph{logical cycle}.

In this work, we quantitatively study the resilience of surface code under the assumption that during ionizing radiation events the device is much noisier than the background error rate for a brief period on the order of the gate time.
We refer to these events as \emph{error bursts} and assume that (1) each error burst occurs within the duration of a single round of syndrome extraction and (2) error bursts are %
rare. Because error bursts are rare, we may regard them as isolated events; that is, typically many rounds of syndrome measurement occur in between successive error bursts. 

We contrast the threshold error rate during error bursts with the threshold for the background error rate. We refer to these two values as the \emph{burst threshold} and the \emph{sustainable threshold}, respectively.
In the absence of measurement errors, all errors are corrected within a single syndrome measurement round, and therefore in that case %
the sustainable threshold and burst threshold are equal.
However, when there are measurement errors, the burst threshold can be higher than the sustainable threshold; while an excessive number of errors occurring during a burst might be intolerable if followed immediately by further rounds with a similarly large error rate, these errors might become correctable if the error rate following the burst quickly settles down to its background value. 
In this work, we investigate the separation between the burst threshold and the sustainable threshold in a simple noise model. Our main result is a relation between the values of these thresholds, depicted in \cref{{fig:circuit_phase_boundary_diagram}}.

\section{Background}

We will analyze a noise model which, although highly idealized, provides a useful caricature of actual errors in realistic devices. Specifically, we consider a depolarizing-channel noise model in which the error rate is enhanced during a burst that lasts for a single syndrome extraction cycle. Here we briefly sketch two possible sources of such error bursts, namely high-energy impact events in superconducting hardware, and global control noise in AMO and other hardware platforms. See Appendix \ref{appendix_sources_of_error_bursts} for further details. 

Ionizing radiation impacting solid-state superconducting devices has been shown to produce highly correlated error bursts~\cite{wilen2021correlated, mcewen2022resolving, mcewen2024resisting, harrington2024synchronous}. The spatial region affected by such an impact event can be larger than the projected size of surface-code logical qubits; hence a spatially uniform error burst is a reasonable representation of the errors induced by these events. While the error bursts found in current devices typically take hundreds to thousands of error correction cycles to abate, mitigation strategies have been proposed that would substantially reduce this timescale~\cite{martinis2021saving, iaia2022phonon}. Therefore studies of our idealized error burst model can illuminate the usefulness of these mitigation strategies. 

Global control is a desirable architectural feature, where a single control system influences many qubits simultaneously. For instance, in AMO systems a single laser might facilitate many gates in parallel, providing very favorable scaling of the control hardware as the number of qubits grows. However, noise in a global control system presents a potential single point of failure, as an aberration in the control signal output can produce errors on all qubits the signal is applied to. %
If substantial deviations in the global control signal are rare and independent over time, they can be accurately described as error bursts. Here, too, analysis of our error burst model is helpful for assessing architectural tradeoffs.

\subsection{Noise model}
If an error burst elevates the gate error rate over many consecutive syndrome extraction cycles, then a logical error is likely to occur unless the error rate during the burst is below the surface code sustainable threshold.
However, if error bursts are brief and rare, then additional resilience may be possible; therefore we study the case where the error burst has a limited temporal extent. The rapid return to the background error rate after the burst might occur because of effective quasiparticle mitigation strategies such as gap engineering or because classical control noise has appreciable high frequency components.

Although in a realistic setting one might expect errors in successive rounds to be correlated, for simplicity we assume there are no such correlations.
Another feature of error bursts is their large spatial extent; we pessimistically assume that the elevated error rate during the burst applies uniformly across the entire surface code block.
It would be useful to extend our analysis to the case where errors in distinct rounds are correlated and error bursts have limited spatial extent, but we have not attempted that more general analysis.

The one-qubit depolarizing noise channel with depolarizing rate \(p\) applies the identity with probability \(1-p\) and one of \(X\), \(Y\), or \(Z\) each with probability \(p/3\).
Likewise, the two-qubit depolarizing noise channel with depolarizing rate \(p\) applies identity with probability \(1-p\) and one of the 15 non-trivial two-qubit Pauli operators each with probability \(p/15\).

We consider the surface code under two noise models, phenomenological noise and circuit-level depolarizing noise.
In both cases, we assign a noise parameter \(p\) to each round of syndrome extraction.

In the phenomenological noise model, bit flip noise (Pauli \(X\) applied with probability \(p\)) is applied to all data qubits with rate \(p\) followed by a round of perfect syndrome extraction.
Finally, all measurement outcomes are flipped independently with probability \(p\). 

In the circuit-level depolarizing noise model, a standard syndrome extraction circuit is executed \cite{FMMC_2012} where for each one (two)-qubit gate in the circuit, the gate is assumed to execute perfectly followed by one (two)-qubit depolarizing noise with rate \(p\).
All measurement outcomes are flipped with probability \(p\).

During a round of syndrome extraction where an error burst occurs, we replace the noise parameter with the burst error rate \(p_B\). In particular, for the circuit-level depolarizing noise model, this implies that all gate operations fail with probability \(p_B\) within the syndrome extraction round where the error burst occurs.

The performance of a stabilizer code under such a noise model can be efficiently simulated using standard open-source tools such as {\it Stim}~\cite{gidney2021stim}.

\subsection{Decoding surface codes}
Surface codes can be efficiently decoded both in theory and in practice \cite{dennis2002topological,Delfosse_2021,higgott2023sparse}.
The efficient decoding of surface codes usually relies on a construction known as the \emph{decoding graph} containing a vertex for every measured stabilizer generator.
For every independent error \(e\), the decoding graph contains an edge incident to the vertices corresponding to the flipped measurement outcomes due to the presence of \(e\).
If the error occurs with probability \(p_e\), the edge is assigned a weight \(\log \frac{1-p_e}{p_e}\).
To ensure that the decoding graph is a well defined graph, Pauli errors are decomposed into a product of Pauli \(X\) and Pauli \(Z\) operators which are assumed to be independent.

\begin{figure}[]
\centering
\def\myscale{1.1}
\tdplotsetmaincoords{80}{120} %
\tdplotsetrotatedcoords{0}{0}{0} %
\centering

\begin{tikzpicture}[scale=\myscale,tdplot_rotated_coords,
                    rotated axis/.style={->,purple,ultra thick},
                    axis/.style={->,blue,ultra thick},
                    blackBall/.style={ball color = black!80},
                    greyBall/.style={ball color = black!20},
                    radiationAlignment/.style={ultra thin, dashed, opacity = 0.6},
                    cube/.style={dashed, ultra thin,fill=red, opacity = 0.4},
                    borderBall/.style={ball color = white,opacity=.25},
                    wave/.style={
                    decorate,decoration={snake,post length=1.4mm,amplitude=1mm,
                    segment length=2mm},red, thick},
                    very thick]
                    \centering
\foreach \x in {0,1,2,3}
   \foreach \y in {-1,0,1,2,3}
      \foreach \z in {0,1,2,3,4,5,6,7}{
           \ifthenelse{  \lengthtest{\x pt < 3pt}  \and \lengthtest{-1pt < \y pt} \and \lengthtest{\y pt < 3pt}}{
             \draw[thick] (\x,\y,\z) -- (\x+1,\y,\z);
             \shade[rotated axis,blackBall] (\x,\y,\z) circle (0.05cm); 
           }{}
           \ifthenelse{  \lengthtest{\z pt = 3pt} \and \lengthtest{\x pt < 3pt}  \and \lengthtest{-1pt < \y pt} \and \lengthtest{\y pt < 3pt} }{
               \draw[ultra thick,red] (\x,\y,\z) -- (\x+1,\y,\z);
               \shade[rotated axis,blackBall] (\x,\y,\z) circle (0.05cm);
           }{}
           \ifthenelse{  \lengthtest{-1pt < \y pt} \and \lengthtest{\y pt < 2pt}  }{
               \draw[thick] (\x,\y,\z) -- (\x,\y+1,\z);
               \shade[rotated axis,blackBall] (\x,\y,\z) circle (0.05cm);
           }{}
           \ifthenelse{  \lengthtest{\y pt = 2pt}  }{
               \draw[thick] (\x,\y,\z) -- (\x,\y+1,\z);
               \shade[rotated axis,blackBall] (\x,\y,\z) circle (0.05cm);
           }{}
           \ifthenelse{  \lengthtest{\y pt = 3pt}  }{
               \shade[rotated axis,greyBall] (\x,\y,\z) circle (0.05cm);
           }{}
           \ifthenelse{  \lengthtest{\y pt = -1pt}  }{
               \draw[thick] (\x,\y,\z) -- (\x,\y+1,\z);
               \shade[rotated axis,greyBall] (\x,\y,\z) circle (0.05cm);
           }{}
           \ifthenelse{  \lengthtest{\z pt = 3pt} \and \lengthtest{-2pt < \y pt} \and \lengthtest{\y pt < 3pt} }{
               \draw[ultra thick,red] (\x,\y,\z) -- (\x,\y+1,\z);
               \shade[rotated axis,blackBall] (\x,\y,\z) circle (0.05cm);
           }{}
           \ifthenelse{  \lengthtest{\z pt < 7pt} \and \lengthtest{-1pt < \y pt} \and \lengthtest{\y pt < 3pt} }{
               \draw[thick] (\x,\y,\z) -- (\x,\y,\z+1);
               \shade[rotated axis,blackBall] (\x,\y,\z) circle (0.05cm);
           }{}
           \ifthenelse{  \lengthtest{\z pt = 3pt} \and \lengthtest{\z pt < 7pt} \and \lengthtest{-1pt < \y pt} \and \lengthtest{\y pt < 3pt} }{
               \draw[ultra thick,red] (\x,\y,\z) -- (\x,\y,\z+1);
               \shade[rotated axis,blackBall] (\x,\y,\z) circle (0.05cm);
           }{}

}

 \draw[axis] (5,0,0) -- (5,0,8) node[anchor=west]{\Large{$T$}};

\draw[->, wave](4.4, -0.5, 3.8) -- (3.7, 0.3, 3.8);
\draw[->, wave](5, -1, 3.4) -- (4.3, -0.2, 3.4);
\draw[->, wave](0, 2.2, 3.4) -- (-0.7, 3, 3.4);
\draw[->, wave](-0.5, 2.7, 3.7) -- (-1.2, 3.5, 3.7);

	\draw[cube] (4,-1,3) -- (4,3,3) -- (4,3,4) -- (4,-1,4) -- cycle;

	\draw[cube] (-1,3,3) -- (4,3,3) -- (4,3,4) -- (-1,3,4) -- cycle;

	\draw[cube] (-1,-1,4) -- (-1,3,4) -- (4,3,4) -- (4,-1,4) -- cycle;

\draw[radiationAlignment](-1, -1, 3) -- (-1, 3, 3);
\draw[radiationAlignment](-1, -1, 3) -- (4, -1, 3);

\draw[radiationAlignment](0, 0, 4) -- (-1, 0, 4);
\draw[radiationAlignment](0, 0, 4) -- (0, -1, 4);
\draw[radiationAlignment](3, 2, 4) -- (3, 3, 4);
\draw[radiationAlignment](3, 2, 4) -- (4, 2, 4);
\draw[radiationAlignment](3, 0, 4) -- (3, -1, 4);
\draw[radiationAlignment](3, 0, 4) -- (4, 0, 4);
\draw[radiationAlignment](0, 2, 4) -- (-1, 2, 4);
\draw[radiationAlignment](0, 2, 4) -- (0, 3, 4);
\draw[radiationAlignment](0, 0, 3) -- (-1, 0, 3);
\draw[radiationAlignment](0, 0, 3) -- (0, -1, 3);
\draw[radiationAlignment](3, 2, 3) -- (3, 3, 3);
\draw[radiationAlignment](3, 2, 3) -- (4, 2, 3);
\draw[radiationAlignment](3, 0, 3) -- (3, -1, 3);
\draw[radiationAlignment](3, 0, 3) -- (4, 0, 3);
\draw[radiationAlignment](0, 2, 3) -- (-1, 2, 3);
\draw[radiationAlignment](0, 2, 3) -- (0, 3, 3);

\end{tikzpicture}
\caption{Decoding graph for a distance-4 surface code.
Black edges correspond to independent error events.
The vertices correspond to the syndrome bits at each spatial position and time. 
A syndrome bit \(v\) is \(1\) if the error set contains an odd number of edges incident to \(v\).
The black edges parallel to the $x{-}y$ plane correspond to errors on the data qubits of the surface code. 
The black edges parallel to the $z$
axis correspond to syndrome measurement errors and are in the error set if the associated measurement is flipped relative to its ideal value.
The decoding graph also corresponds to the noise model of the code: To every edge, we can associate an independent random indicator variable that is \(1\) if the edge is contained in the error set and \(0\) otherwise.
The surface code undergoes $T$ consecutive rounds of syndrome extraction with an error burst occurring at round $T/2$. 
The edges in that cycle are marked in red, indicating that the indicator variables are 1 with enhanced probability during that cycle.
    \label{fig:decoding-graph-ionizing-radiation}}
\end{figure}
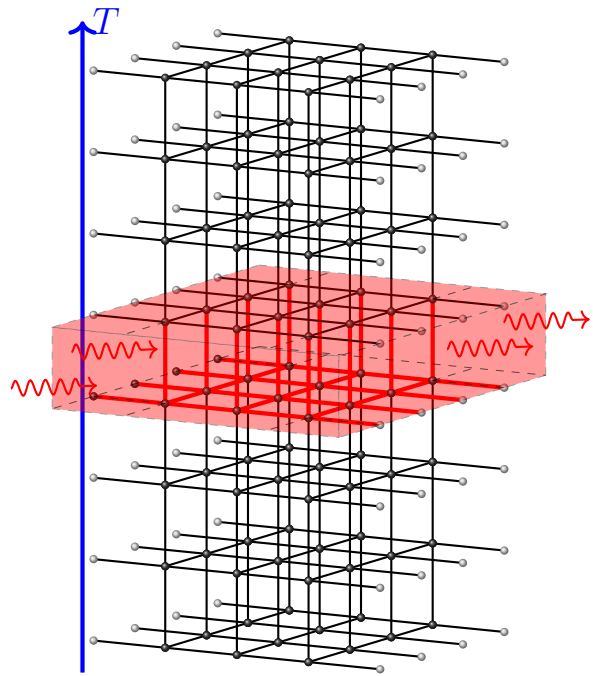

In all numerics, we assume knowledge of the time at which the error burst occurs and update the edge weights appropriately.
This decoding graph for a distance-4 surface code under the phenomenological noise model is illustrated in \cref{fig:decoding-graph-ionizing-radiation}.
However, we find that it is easy to infer the presence of an error burst event by considering the Hamming weight of the change in syndrome in each round: An error burst with burst error rate much higher than background error rate can be easily distinguished from the background error rate.
Appendix~\ref{appendix:error-burst-detection} shows an example maximum-likelihood estimate of identifying whether the noise in a syndrome extraction round is drawn from a distribution with one of two error rates.
In a real system, the distribution of the global noise parameter will have support on many values. Estimation of the noise is still possible, but it may be the case that a Bayesian framework with carefully selected prior would be more suitable.

In this work, we use an open source implementation of the minimum-weight perfect-matching (MWPM) decoder \cite{pymatchingv2} on the weighted decoding graph derived from the noise model \cite{gidney2021stim}.

\subsection{Threshold Experiment}
We consider the rotated surface code \cite{horsman2012surface} subjected to a memory experiment of duration \(T = 2d\) syndrome extraction cycles where a single error burst occurs at time \(T/2\) and $d$ is the surface code distance.
Using Monte Carlo sampling, we estimate the threshold of the surface code under phenomenological and circuit-level depolarizing noise with rate \(p\) including an error burst at time \(T/2\) with rate \(p_B\). 
The memory experiment consists of sampling the noise via Monte Carlo sampling with a final round of perfect readout in the \(Z\) basis.
A failure is declared if the eigenvalue of the logical \(Z\) operator has been flipped by the error and correction.

\begin{figure}
\includegraphics[width=\columnwidth]{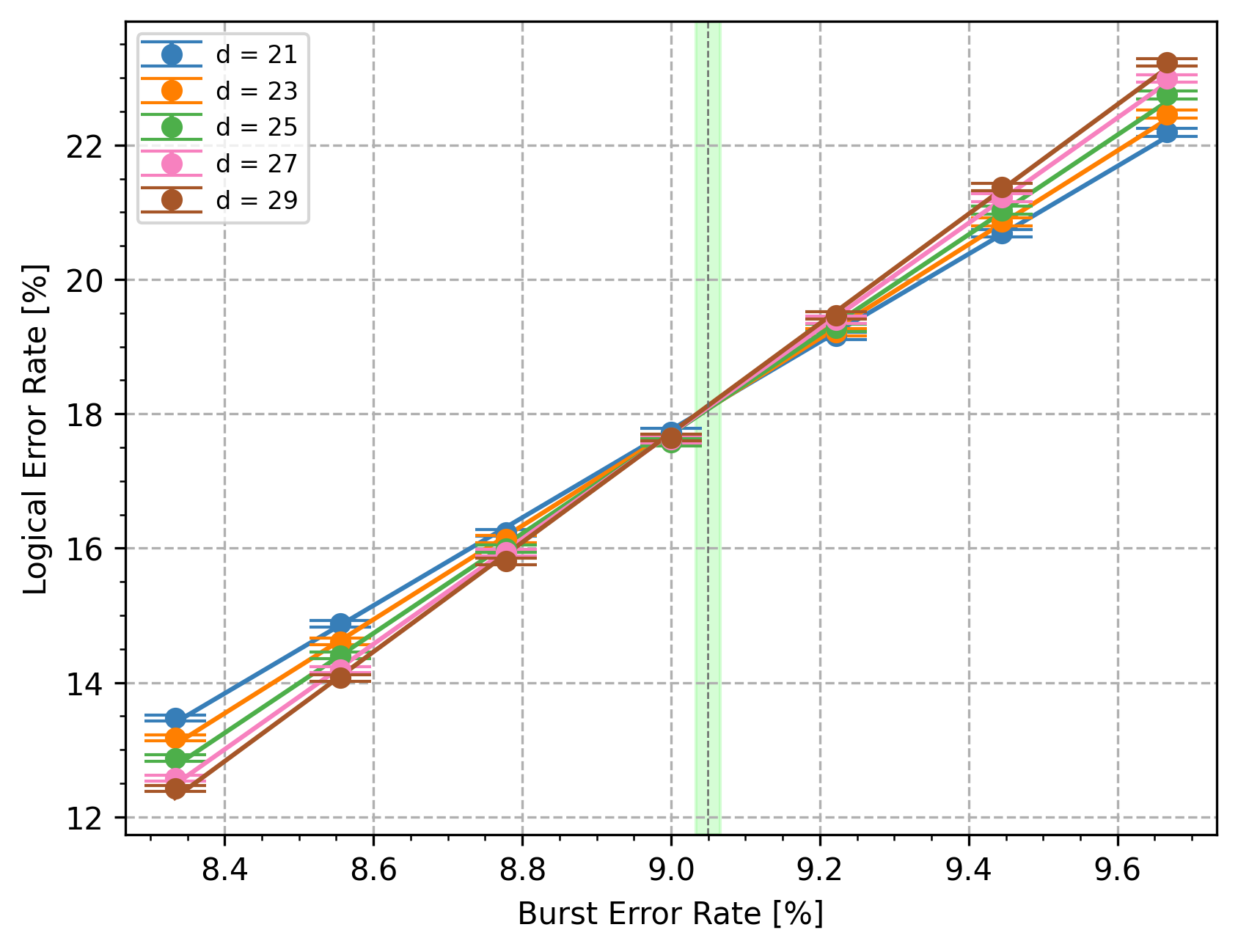}%
\caption{\label{fig:phenom_threshold_plot}
Sweep of \(p_B\) indicating a threshold for phenomenological noise with background error rate $p = 2.0\%$.
The error burst threshold \(p_B^* = 9.050(16)\% \) is indicated by the grey vertical line with green error bars.
}
\end{figure}

Below the threshold, increasing the code distance exponentially suppresses logical errors whereas above the threshold increasing code distance increases the failure rate of the memory.
For each code distance, we run a sweep of the memory experiment with different values of the background or burst error rate to observe the characteristic crossing of the logical failure rates corresponding to the threshold (\cref{fig:phenom_threshold_plot}).
To reduce finite size effects, in the different limits, low measurement failure rate and low burst error rate, we sweep the burst error rate and background error rates, respectively.
The threshold estimation method is described in more detail in Appendix~\ref{sec:estimating_error_burst}.

A single error burst is modeled in order to have a well defined threshold.
However, the conclusions should hold as long as the surface code distance is much smaller than the average time between error bursts: In other words, the ``correlation length'' of the decoding problem under independent noise is on the order of the surface code distance \cite{dennis2002topological}.
In typical devices, the average time between error bursts is more than 6 orders of magnitude greater than the gate time \cite{google2023suppressing}, so we are in this regime for all practical purposes. We observe similar results in the setting of depolarizing circuit level noise.

 Our main results are shown in figs.~\ref{fig:phenom_phase_boundary_diagram} and \ref{fig:circuit_phase_boundary_diagram}. We find that the memory is below threshold even for relatively large values of both the background error rate and burst error rate.
This is the regime of practical interest for near-term devices.
We note that achieving good resilience to both noise sources simultaneously is non-trivial since the presence of one noise source weakens resilience to the other noise source. Our results for the circuit level depolarizing noise model suggest that, based on estimates in \cite{martinis2021saving}, hardware mitigation and standard fault tolerance techniques alone may suffice to protect against ionizing radiation and global control noise.

\section{Results}

\subsection{Thresholds}
\label{sec:phase_diagram}

\begin{figure}
\includegraphics[width=\columnwidth]{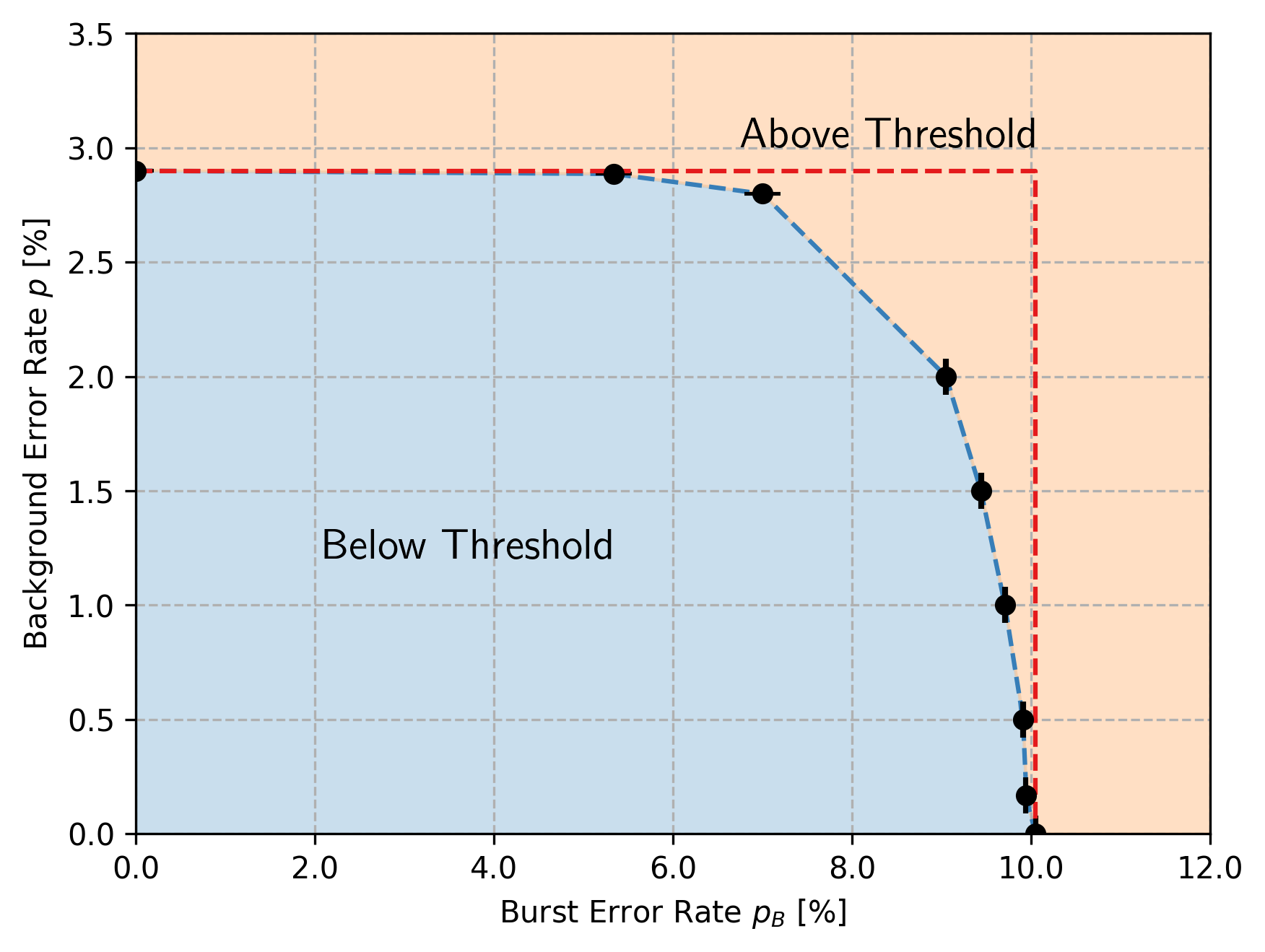}%
\caption{\label{fig:phenom_phase_boundary_diagram}
Phase diagram of the surface code under phenomenological noise.
The $x$-axis and $y$-axis indicate burst and background error rates respectively.
The light tan area indicates a noise rate that is above threshold while the dark blue area indicates a noise rate below threshold.
Data points are connected by a blue dashed line as a guide to the eye.
Limiting behavior in the \(p\to 0\) and \(p_B \to 0\) limits is drawn as a red box indicating upper bounds on the threshold error rates.
}
\end{figure}

We begin by studying the threshold behavior of surface codes in a memory experiment of duration \(T = 2d\) where a single error burst occurs at time \(T/2\) and $d$ is the surface code distance. We verified that similar threshold estimates are obtained when the duration is longer than $T=2d$.

The statistical mechanics mapping for such an error model ~\cite{dennis2002topological,chubb2021statistical} corresponds to a 3D random-bond Ising model with increased disorder along a 2D surface immersed in the bulk.
Related models have been considered previously in the context of noisy lattice surgery ~\cite{ramette2023fault}, where the elevated error rate occurs at the interface between surface code blocks, and quantum communication~\cite{fowler2010surface}, where the elevated error rate characterizes noisy Bell pairs shared by nodes in a quantum network.
We present the phase diagram under phenomenological noise in \cref{fig:phenom_phase_boundary_diagram}.
Notably, we find that the surface code is stable to even large error bursts at finite background error rates which {\it a priori} is not a given.
That is, the presence of additional errors accumulating at a small rate does not significantly harm the ability to recover from a large amount of pre-existing errors.
Quantitatively, under phenomenological noise, the presence of an error burst with rate \(p_B=7\%\), only depresses the sustainable threshold from \(\approx 3\%\) to \(\approx 2.7\%\).

A question of practical relevance is the setting where error bursts occur at a fixed rate \(\Gamma\). After each error burst, the residual population of qubit errors settles back down to the background value after about $d$ syndrome measurement cycles. Thus, for \(d \ll 1/\Gamma\), the \emph{rare event regime}, the actual resilience to error bursts is captured well by our idealized setting in which only a single error burst is considered. But in the opposite regime where \(d \gg 1/\Gamma\), a second error burst is likely to occur before the effects of the previous error burst have fully relaxed. 
Indeed we expect that in the thermodynamic limit, \(d \to \infty\), \(\Gamma = \mathrm{const.}\), there may not be a stable phase with \(p_B > p\).
However, \(d\) is only required to increase poly-logarithmically with the size of the computation, so a polynomial increase in \(1/\Gamma\) leads to an exponential increase in the size of circuit that can be executed while remaining in the rare event regime.

\begin{figure}
\includegraphics[width=\columnwidth]{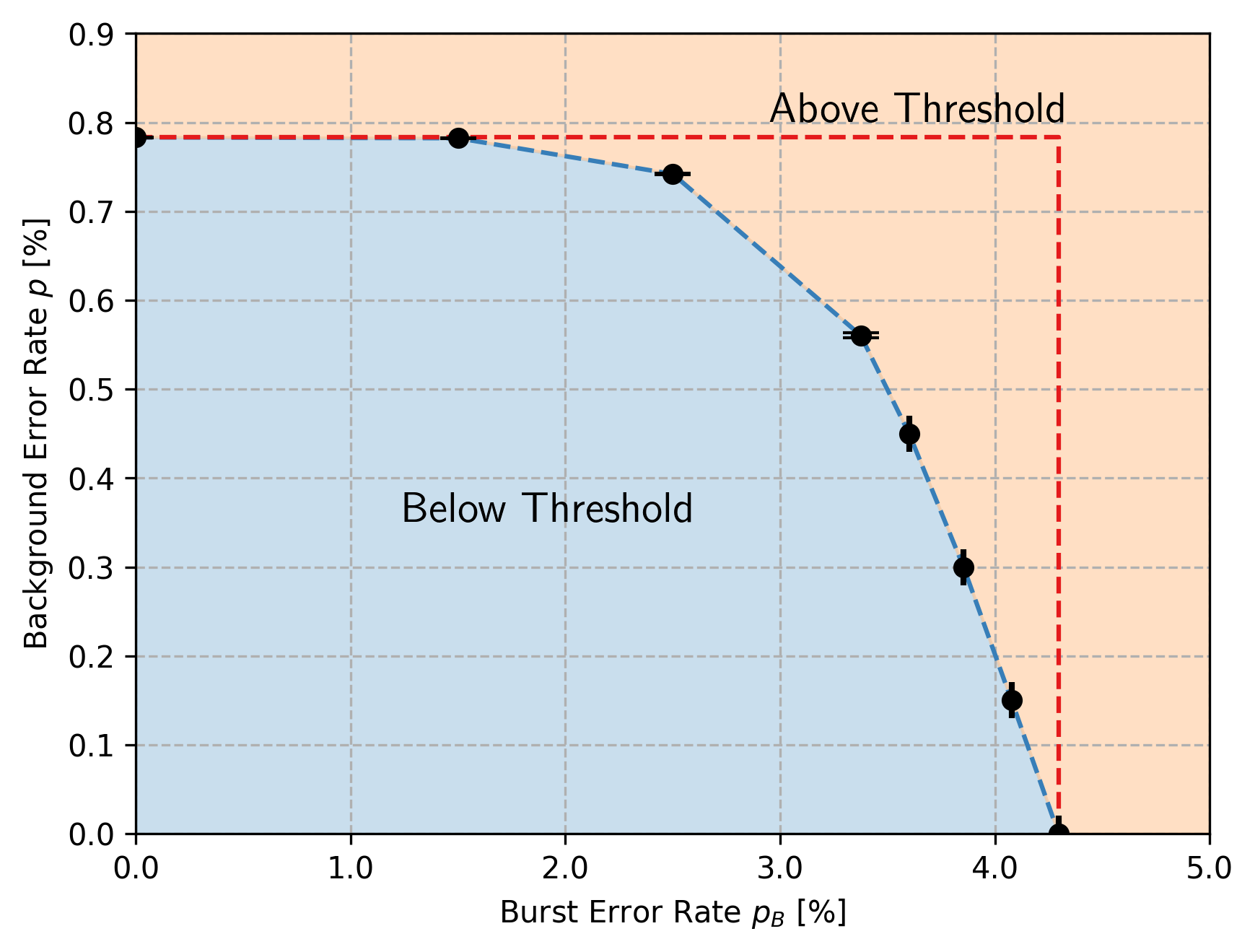}%
\caption{\label{fig:circuit_phase_boundary_diagram}Phase diagram of the surface code under circuit-level depolarizing noise.
The $x$-axis and $y$-axis indicate burst and background error rates respectively.
The light tan area indicates a noise rate that is above threshold while the dark blue area indicates a noise rate below threshold.
Data points are connected by a blue dashed line as a guide to the eye.
Limiting behavior in the \(p\to 0\) and \(p_B \to 0\) limits is drawn as a red box indicating upper bounds on the threshold error rates. }
\end{figure}

By performing a similar study for circuit-level depolarizing noise, we obtain the phase diagram in \cref{fig:circuit_phase_boundary_diagram}; here too we find that the threshold background error rate is relatively stable when the burst error rate is not too high. 
Notably, though, the threshold burst error rate has a stronger dependence on the background error rate than for the phenomenological noise case.
A reasonable hypothesis is that this distinct behavior exhibited by the two phase diagrams occurs because in the circuit-based noise model a larger portion of the errors introduced during the burst remain extant in the following rounds. Thus the background error rate must be reduced accordingly to prevent the error population from surpassing what the decoder can handle. 

\begin{figure}
\includegraphics[width=\columnwidth]{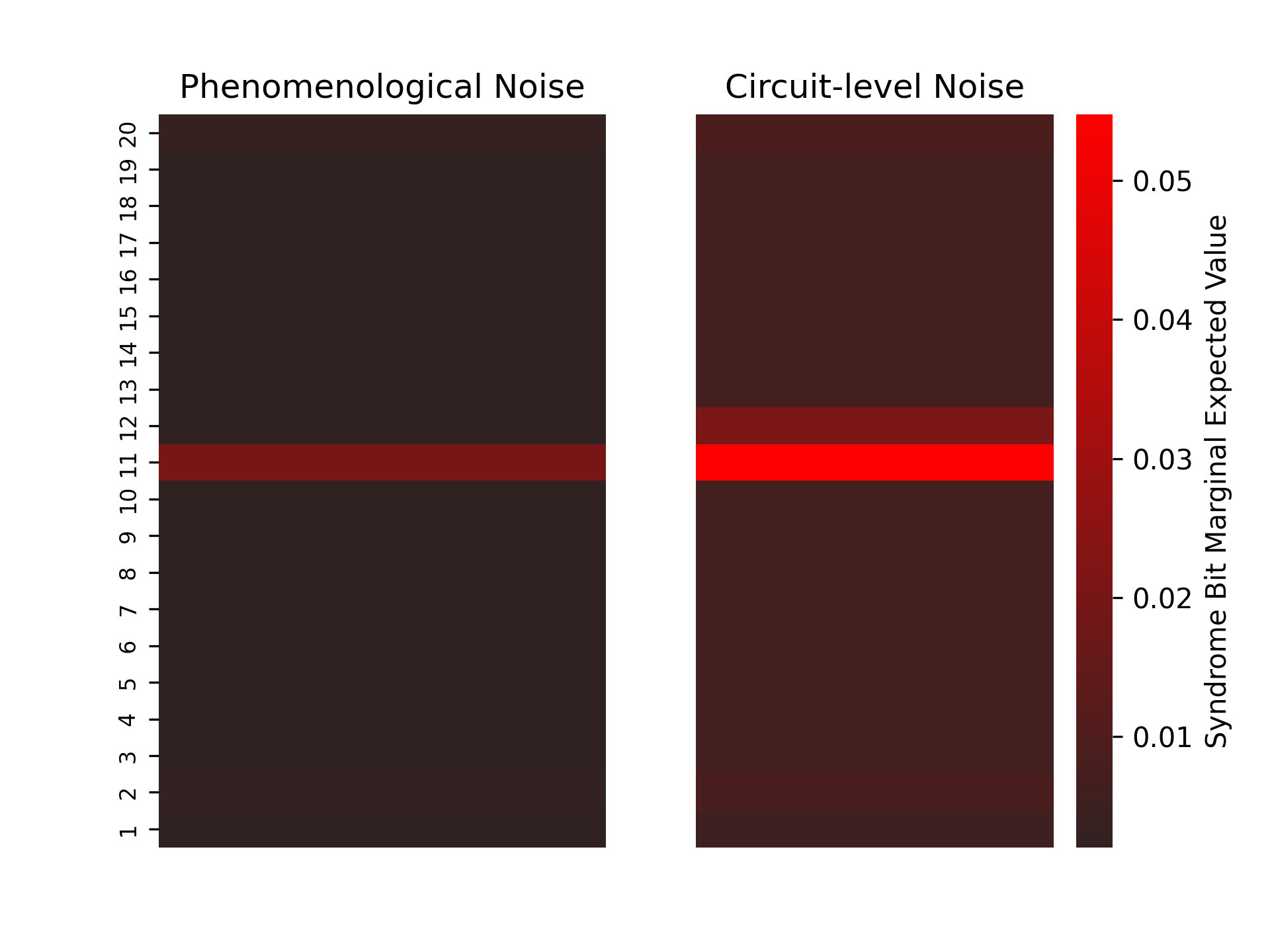}%
\caption{\label{fig:density_of_errors}
Marginal expected value of syndrome bits averaged over each time slice in  a surface-code memory experiment with an error burst. The marginal syndrome bit value is a proxy for the density of errors in a syndrome extraction cycle. Here the code distance is $d=3$, and 20 rounds of syndrome extraction are shown, with an error burst in the 11\textsuperscript{th} round. The background error rate is $p = 0.1\%$ and the burst error rate is $p_B = 1\%$. The effect of the error burst is confined to a single syndrome extraction cycle for the phenomenological noise model; however, it extends into the following syndrome extraction cycle for circuit-level noise because some errors occurring during the burst are not detected until the following round.
}
\end{figure}

To validate this explanation, we evaluate a proxy for the density of errors by computing the marginal expected value of syndrome measurements averaged over a spatial slice, as shown in \cref{fig:density_of_errors}. 
The effect of the burst on the error population is clearly confined to a single time slice in the phenomenological noise model, but the error density remains elevated for two consecutive time slices in the circuit-level depolarizing noise model. 
This effect arises because some of the errors produced in the burst are not detected until the following syndrome extraction cycle. These surviving errors from the burst are augmented by additional errors which occur at the background error rate in the following cycle, explaining the stronger combined effects of burst errors and background errors revealed by the phase diagram.

\subsection{Teraquop footprint}
In order to quantitatively estimate the effect of error bursts on the amount of resources required to execute large circuits under circuit-level depolarizing noise, we estimate the teraquop footprint \cite{gidney2021fault}.
Roughly, the teraquop footprint is the space overhead required to execute a circuit of size \(10^{12}\) such that the probability of success is roughly \(1/e\).
This is achieved by a logical failure rate of \(10^{-12}\).

\subsubsection{Logical error rate}

To separately extract the background logical failure rate and logical failure rate due to error bursts, we run two memory experiments A and B.
Experiment A is a standard \(D\) logical-cycle memory experiment using a distance-\(d\) surface code with background error rate \(p\) and background logical error rate \(q_d(p)\).
In the regime \(D \gg 1\) and assuming the failure for each logical cycle is i.i.d. distributed, 
the failure probability \(\tilde{p}_{d,A}(p,D)\) of the memory experiment is related to the failure probability per logical cycle as %
\begin{align}\label{eq:ansatz_p_L}
   2 \tilde{p}_{d,A}(p,D) \approx 
    1-[1-q_d(p)]^D.
\end{align}
The error rate approaches \(1/2\) as \(D \to \infty\), because a correction selected uniformly at random has probability \(1/2\) of being in the same equivalence class as the error.
In memory experiment B, we use the same noise model except we inject an error burst with error rate \(p_B\) in the $\left(\frac{D}{2}\right)$\textsuperscript{th} logical cycle.
If the logical cycle containing the error burst fails with probability \(q_{d,B}(p_B, p)\), then we expect the failure probability \(\tilde{p}_{d,B}(p_B, p, D)\) in memory experiment B to be  %
\begin{align}\label{eq:ansatz_p_LB}
    2\tilde{p}_{d,B}(p,D) \approx1- [1-q_{d,B}(p_B, p)][1-q_d(p)]^{D-1}.
\end{align}
In other words, we attribute all excess logical failures in experiment B to the presence of the error burst, allowing separate determinations of the background and burst logical error rates.
Using \cref{eq:ansatz_p_L} and \cref{eq:ansatz_p_LB}, we estimate \(q_d(p)\) and \(q_{d,B}(p_B,p)\), finding that these quantities approach constant values independent of $D$, denoted \(\tilde{q}_d(p)\) and \(\tilde{q}_{d,B}(p_B,p)\), when $D$ is large.
A more extended discussion of finite size effects from finite duration memory experiments can be found in \cite[Appendix D]{pattison2021improved}. %
We extrapolate \(\tilde{q}_d(p)\) and \(\tilde{q}_{d,B}(p_B,p)\) to large $d$, and correspondingly small logical failure rates, in order to estimate the teraquop footprint.
A memory experiment with a moderate error burst rate is illustrated in in \cref{fig:mem_expt_B}.
The low error rate extrapolation is explained in greater detail in Appendix~\ref{sec:low-error-rate-extrapolation}.

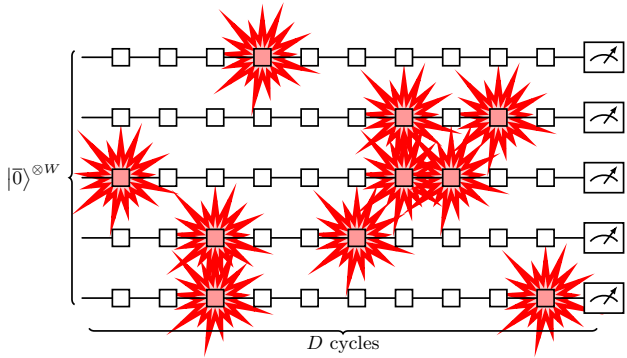
\begin{figure}
\newcommand{\noisygate}{\gate[style={fill=red!40}]{}\gategroup[wires=1, steps=1, style={noisy},background,label style={label position=below,anchor=north}]{{}}}
\centering
\resizebox{\linewidth}{!}{		
\tikzset{
noisy/.style={starburst,fill=red!20,draw=red,line
width=2pt,inner xsep=-2pt,inner ysep=-5pt}
}
\begin{quantikz}[execute at end picture={
    \tikzmath{
      \braceoffset = -0.5;
    }
    \draw
    (\tikzcdmatrixname-5-1) ++(0,\braceoffset)  node(A) {}
    (\tikzcdmatrixname-5-12) ++(0,\braceoffset) node(B) {};
    \draw[decorate,decoration={brace,mirror},thick] (A) -- (B) node[midway,anchor=north]{\(D\) cycles};
}]
\lstick[5]{$\ket{\overline{0}}^{\otimes W}$} & \gate{}     & \gate{} & \gate{}    & \noisygate & \gate{} & \gate{}    & \gate{}    & \gate{}    & \gate{}    & \gate{}    & \meter{} \\
                                           & \gate{}     & \gate{} & \gate{}    & \gate{}    & \gate{} & \gate{}    & \noisygate & \gate{}    & \noisygate & \gate{}    & \meter{}\\
                                           & \noisygate  & \gate{} & \gate{}    & \gate{}    & \gate{} & \gate{}    & \noisygate & \noisygate & \gate{}    & \gate{}    & \meter{}\\
                                           & \gate{}     & \gate{} & \noisygate & \gate{}    & \gate{} & \noisygate & \gate{}    & \gate{}    & \gate{}    & \gate{}    & \meter{}\\
                                           & \gate{}     & \gate{} & \noisygate & \gate{}    & \gate{} & \gate{}    & \gate{}    & \gate{}    & \gate{}    & \noisygate & \meter{}
\end{quantikz}
}
  \caption{
      Logical quantum circuit for memory experiment B with $W$ logical qubits and $D$ logical cycles such that $W \cdot D = 10^{12}$. 
      Uncolored blocks are logical cycles that only suffer from the background error rate and fail with probability $\tilde{q}_{d}(p)$. Red blocks are logical cycles in which an error burst occurs; these fail with probability $\tilde{q}_{d,B}\left(p_B, p\right)$. \label{fig:mem_expt_B}
 }
\end{figure}

\subsubsection{Failure model}
If the average number of logical cycles between error bursts is \(\tau \ge 1\), the probability $P_L(p_B,p)$ of a logical failure per logical cycle-qubit is approximately
\begin{align}\label{eq:encoded-fail-prob}
    P_L(p_B,p) \approx \frac{1}{\tau} \cdot \tilde{q}_{d,B}(p_B, p) + \left(1 - \frac{1}{\tau}\right)\cdot \tilde{q}_d(p).
\end{align}
We assume a surface code of distance \(d\) requires a total of \(2d^2\) qubits including ancilla qubits~\footnote{A rotated surface code only requires \(2d^2-1\) however a regular tiling of surface codes will require an extra qubit.}.
We define the teraquop footprint as the total number of physical qubits (including ancilla qubits) per logical qubit of the minimum distance surface code that achieves a logical error rate of \(10^{-12}\).

\begin{figure}
    \centering
    \includegraphics[width=\columnwidth]{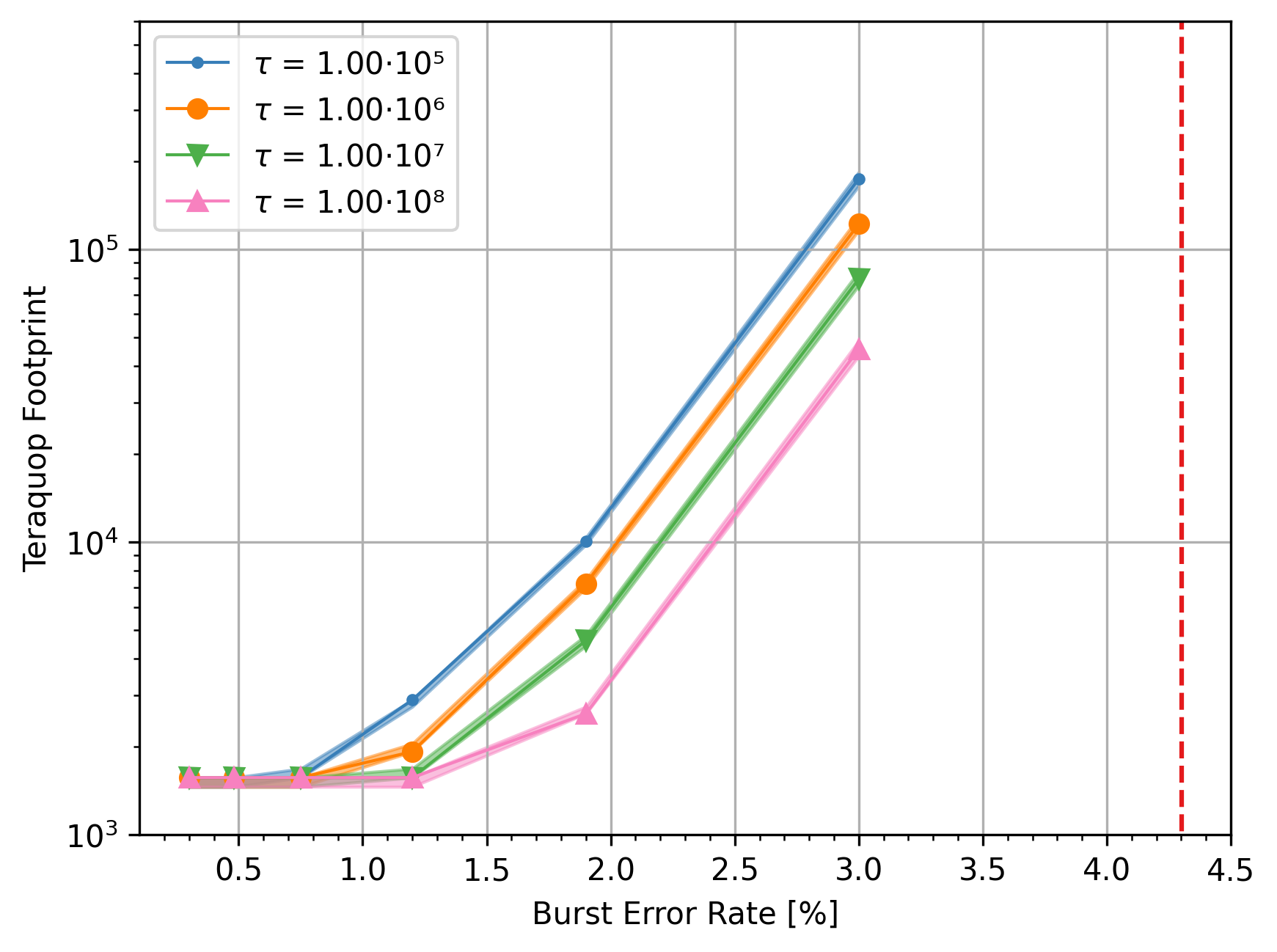}
    \caption{Teraquop footprint for background error rate \(p = 0.1\%\) as a function of the burst error rate $p_B$ for various values of the mean time $\tau$ between error bursts. The burst error threshold in the limit \(p\to 0\) is indicated as a vertical dashed line.}
    \label{fig:tau_teraquop_footprint_0001}
\end{figure}

If a logical cycle takes about \SI{10}{\micro\second}, then an error burst occurring once per minute corresponds roughly to \(\tau \approx 10^7\).
In \cref{fig:tau_teraquop_footprint_0001}, we show the teraquop footprint as a function of the mean time between error bursts \(\tau\) and the error burst rate when the background error rate is \(p=10^{-3}\).
We provide fit parameters used for the teraquop footprint estimate in \cref{tab:fit_params} in the appendix.

Notably, an error burst rate %
an order of magnitude above the background error rate does not appreciably affect the teraquop footprint.
However, performance quickly degrades beyond this point.
At low background error rates, the teraquop footprint is also relatively insensitive to further reduction in the background error rate. 

\section{Discussion and Conclusion}
We have shown that surface codes are resilient to rare ``error bursts'' in which the physical error rate is briefly enhanced. 
Such error bursts have many potential sources and become relevant once the frequency of burst events becomes comparable to the inverse of the duration of the longest computation wall time we wish to consider (\textit{e.g.}, days to weeks).
Because such events are rare they may escape notice in experiments with insufficient runtime. 
Superconducting qubit devices stand apart in that error bursts originating from ionizing radiation have been directly detected.
But in a variety of quantum platforms engineering constraints favor control signals shared by many qubits. 
This engineering convenience carries the risk of potential error burst failure modes which should be understood and mitigated.

With appropriate mitigation methods, the burst event can be brief; furthermore, the burst error rate, though higher than the background error rate, may not be so high as to render quantum error correction ineffective. We find that if the mean time between error bursts is much larger than the duration of a full logical cycle, and if the burst error rate is below the burst threshold, then the overhead cost of error correction is rather insensitive to the burst error rate. 
For reasonable parameters (\SI{10}{\micro\second} logical cycle, \SI{1}{\per\min} mean time between bursts, background error rate $p=.1\%$),  we find that if the burst error rate is  \(15\) times larger than the background error rate, then the teraquop footprint increases by less than a factor of 2 compared to the case without any burst errors.

In our computations, we assumed that the decoder knows the time step in which a burst error occurs. This assumption is reasonable because the elevated error rate during the burst is easily detected in the syndrome measurement history. Furthermore, we expect the decoder to perform nearly as well even when prior information about the time of the burst is not provided. 

Our studies have been limited to the surface code, but we anticipate that qualitatively similar results apply when using other families of quantum low-density parity-check codes. Our conclusions bolster the hope that, given appropriate mitigation strategies that diminish the temporal extent and strength of error bursts, deep quantum circuits with valuable practical applications can be executed reliably on fault-tolerant quantum computers.

\section{Acknowledgments}
We thank Steve Flammia, Hengyun Zhou, Dolev Bluvstein, Pedro Sales Rodriguez, and Adam Shaw for valuable conversations. 
SJST acknowledges funding from Caltech's Summer Undergraduate Research Fellowship (SURF) and QuICS's Lanczos Graduate Fellowship.
CAP acknowledges funding from the Air Force Office of Scientific Research (AFOSR) FA9550-19-1-0360 and U.S. Department of Energy Office of Science, DE-SC0020290.
JP acknowledges support from the U.S. Department of Energy Office of Science, Office of Advanced Scientific Computing Research (DE-NA0003525, DE-SC0020290), the U.S. Department of Energy, Office of Science, National Quantum Information Science Research Centers, Quantum Systems Accelerator, and the National Science Foundation (PHY-1733907). 
The Institute for Quantum Information and Matter is an NSF Physics Frontiers Center.

\bibliography{apssamp}%

\appendix
\onecolumngrid

\subsection{Sources of error bursts}\label{appendix_sources_of_error_bursts}
Here, we discuss in more detail two motivating cases of error sources in specific hardware implementations that are modeled well as error bursts, namely quasiparticle bursts in superconducting qubit platforms, and global control noise in AMO and other platforms.

\subsubsection{Quasiparticle bursts from high-energy impact events}
When ionizing radiation (for example a cosmic ray muon or gamma radiation) interacts with the chip substrate, large amounts of energy are deposited (100keV-1MeV), which quickly radiates outwards in the form of high-energy phonons~\cite{wilen2021correlated, mcewen2022resolving}. When these phonons impinge on the superconducting qubit material, they break Cooper pairs and produce Bogoliubov quasiparticles. The quasiparticles can diffuse through the qubit and tunnel through the Josephson junction, where they can absorb the qubit energy. This tunneling process severely limits the $T_1$ relaxation time of the qubits, leading to a substantial increase in the error rate. Due to the rapid ballistic transport of high-energy phonons through the substrate, these elevated error rates are experienced over a large area, comparable to the size of current superconducting qubit arrays. 

The error rate returns to normal when the quasiparticle population density relaxes to its background value. Quasiparticles disappear as they recombine to form Cooper pairs, releasing phonons.
These phonons escape slowly off the device, producing long recovery timescales between 100s of microseconds~\cite{wilen2021correlated} and 10s of milliseconds~\cite{mcewen2022resolving, mcewen2024resisting, harrington2024synchronous}. Recombination requires two quasiparticles to interact, and such interactions become less frequent as the population thins out; therefore the recombination rate declines as the density is reduced, and as a result the quasiparticle density decreases only polynomially with time~\cite{lenander2011measurement}.

Various mitigation strategies have been proposed~\cite{martinis2021saving} and implemented.
Phonon trapping~\cite{iaia2022phonon} aims to down-convert phonons to a frequency lower than the qubit superconducting gap, preventing quasiparticle generation in the qubits.
Quasiparticle trapping~\cite{thorbeck2023tls} aims to concentrate the quasiparticle populations away from the qubit Josephson junction, limiting the errors induced by the elevated quasiparticle population and aiding recombination.
These mitigation strategies can enhance the speed of recovery to normal performance to the extent that only a single round of error correction is affected~\cite{thorbeck2023tls}.
While the spatial extent of the error burst is not infinite, and may be significantly improved by these mitigation strategies, a uniform error burst over all qubits represents a worse case for QEC performance.

\subsubsection{Global control noise}
\begin{figure}
    \centering
    \includegraphics[width=0.7\columnwidth]{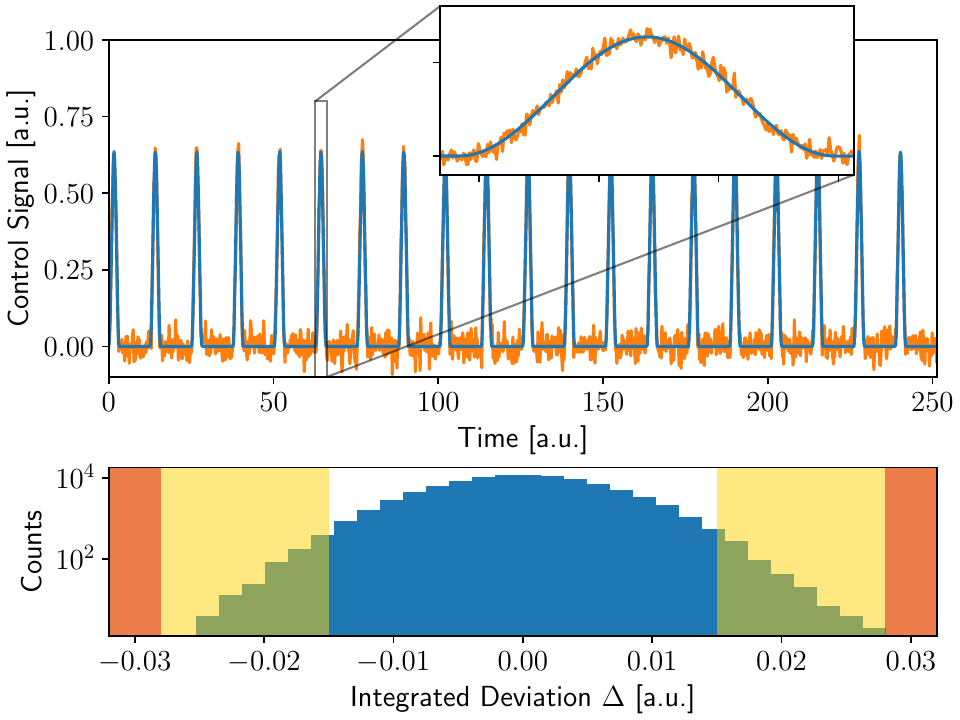}
    \caption{\label{fig:param-trace}
    (Top) Illustration of hypothetical global control parameter \(g\) with periodic pulses (e.g. envelope of laser amplitude modulation).
    The ideal value is indicated in blue while orange indicates a trace with white noise added (independent in time, marginals distributed as \(\mathcal{N}\left(0,\sigma^2=9\cdot 10^{-4}\right)\)).
    This is the ``best'' case scenario as such noise is well behaved and has low tail weight.
    (Bottom) Histogram of the deviation \(\Delta\) of the integrated signal within a pulse from the ideal value, taken over \(10^5\) pulses.
    Note the logarithmic Y-axis.
    In the discussion of the behavior of \(\Delta\), the region \(|\Delta| \ge \Delta_*\) is illustrated as a yellow shaded region, while a ``never exceed'' value of \(\Delta\) is illustrated as a red shaded region.
    Due to the unbounded nature of the noise, arbitrarily large deviations can be encountered.
    }
    \label{fig:parameter-trace}
\end{figure}

Implementing gates in quantum computation experiments requires precise time-variation in Hamiltonian parameters.
To ease engineering challenges, it is frequently the case that a single device is utilized in some step of the control electronics for the control of many qubits (e.g. shared microwave sources, arbitrary waveform generators, lasers).
Such sharing of resources makes the system \emph{formally} not fault-tolerant, but it may be reliable enough for practical purposes.
Since the time-varying Hamiltonian control parameter is a real number (analog), the deviation in the control system from the ideal output is also characterized by a real number which is time varying and in principle unbounded.

This raises the question of the correct figure-of-merit for reliability.
A failure corresponds to a large deviation in a control signal such that the executed gate is very far from the ideal gate.
If such a failure occurs in a component used for global control, the resulting error burst may be fatal for a fault-tolerant quantum memory~\footnote{An extremely large global coherent error deviates from our definition of an error burst as elevated independent noise, but moderately large coherent errors are approximately captured by our definition. }.

Let $g$ denote a global control parameter that controls the execution of many gates in parallel in an error-corrected quantum memory, and let \(\Delta \in \mathbb{R}\) denote the deviation of $g$ from its ideal value. Naively, to ensure good performance, one might require that $|\Delta|\le \Delta_*$ be satisfied with high probability in each syndrome extraction step for some suitably small $\Delta_*$. However, if the memory is sufficiently resilient to brief error bursts, then occasional large excursions of $g$ with $|\Delta|$ significantly larger than $\Delta_*$ (but not too much larger) can also be tolerated. This weaker control requirement is easier to achieve in practice. We illustrate an example trace of \(g\) in  \cref{fig:parameter-trace} and the distribution of integrated deviation for each control pulse.

For illustrative purposes, we proceed to describe how rare error bursts might arise from global control in an optical tweezer array, and to note that other quantum computing architectures have similar vulnerabilities.

In platforms based on atomic, molecular, and optical (AMO) physics, global control is often very natural due to the resource-intensive nature of coherent light sources. Here, we describe the control method of a particular neutral atom quantum computing experiment reported on in \cite{evered2023high}, but one can expect similar error mechanisms to be present in other experiments.

Evered {\it et al.}~\cite{evered2023high} report on a high-fidelity (\(\approx 99.5\%\)) entangling gate in a quantum computing experiment based on neutral Rubidium-87 atoms trapped in optical tweezers.
The qubit computational basis is formed by the two hyperfine states \(\ket{0}=\ket{F=0, m_F=0}\) and \(\ket{1}=\ket{F=0, m_F=2}\).
Entanglement of two nearby atoms can be realized by transferring the \(\ket{1}\) state to a state with high principle quantum number known as a Rydberg state.
These excited states have strong interactions such that, after returning to the qubit subspace, the state \(\ket{11}\) accumulates a phase while \(\ket{00}\), \(\ket{10}\), and \(\ket{01}\) do not.

In \cite{evered2023high}, the physical entangling gate is applied to all atoms in a pre-determined region by shining \SI{420}{\nano\meter} and \SI{1013}{\nano\meter} laser light modulated by an acousto-optic modulator over the course of approximately \SI{200}{\nano\second}.
This corresponds to a two-photon transition to an \(n=53\) Rydberg state via an intermediate state from which the laser light is detuned.
Such a gate scheme naturally lends itself to scaling up, because the control degrees of freedom are mostly independent of the number of atoms in the gate region.
However, variation in the pulse shape may cause a large global error on all atoms in the gate zone.
For example, intensity noise can change the two-photon Rabi frequency, leading to over/under-rotation and a final state that is not completely in the qubit subspace (i.e. a coherent leakage error).
For moderate values of the rotation error, this deviation from ideal control results in an error burst affecting many qubits.

Because such error bursts are rare events, it is difficult to detect them directly except by running an error correction experiment or some other specialized experiment that gathers a large amount of data. We are not aware of any dedicated search for error bursts in AMO platforms performed to date, but imminent experiments running full-scale quantum error correction are likely to provide instructive quantitative information about the nature and rate of such correlated errors.

A fundamental question is that of mitigation strategies after identifying a global noise source.
As global noise sources are highly setup-specific, the mitigation strategy will also be setup-specific.
Our numerics provide a guideline for how much mitigation is necessary.
Using the previous example of laser intensity noise on a global gate, closed loop control may suffice for low frequency (slowly-varying) noise.
For high-frequency noise, a time-multiplexing strategy where gates are only done on subsets of atoms at a time may suffice.
One could also take advantage of the long coherence time of atomic platforms and ``pause'' the computation until the noise returns to an acceptable range.

Correlated gate errors resulting from global control are not unique to AMO platforms; in particular spin-qubit architectures frequently rely on global control.
For instance, in nitrogen-vacancy center qubits in diamonds, architectures commonly feature global optical and microwave control signals \cite{neumann_quantum_2010, bradley_ten-qubit_2019, abobeih_fault-tolerant_2022}.
Similarly, in electron spin qubits, global microwave or DC electrical control is a common architectural feature~\cite{zwanenburg_silicon_2013, hill_surface_2015, veldhorst_silicon_2017}. 
In all of these cases, classical control noise can induce fidelity correlations among gates performed in parallel, and rare large deviations of control signals from their ideal values could cause error bursts.

\subsection{Error burst detection}\label{appendix:error-burst-detection}
Here, we show that a priori knowledge of the time at which an error burst occurs is not necessary.
The presence of an error burst can be inferred directly from the syndrome.
We will prove bounds on the probability of a false positive or false negative in a toy model using a maximum likelihood estimator.
For simplicity, consider a subset of vertices \(V\) in a decoding graph contained in a single time slice such that each vertex has \(w\) fault locations incident to it and no two vertices in \(V\) are adjacent.
Then, suppose that each fault location contains an error with the background error probability \(p\) while during an error burst it contains an error with probability \(p_B > p\) less than \(1/2\).

The assumption of uniform noise and degree can be straightforwardly relaxed, and the subset of vertices can be found by exploiting structure in the decoding graph.
If the decoding graph is sparse, one can also find a greedy approximate solution to the maximum independent set problem \cite{halldorsson1994greed}.

We wish to determine whether, at an arbitrary time \(t\), there was an error burst.
Define \(p_1 = \sum_{i=1}^{\lceil w/2 \rceil} \binom{w}{2i-1} p^{2i-1} (1-p)^{w-2i+1}\), \(p_2 = \sum_{i=1}^{\lceil w/2 \rceil} \binom{w}{2i-1} p_B^{2i-1} (1-p_B)^{w-2i+1}\) to be the marginal probabilities that an odd number of faulty edges are incident to  a fixed vertex of \(V\).
Let \(q\) taking values in \(\{p_1,p_2\}\) be the random variable corresponding to the marginal flip probability of syndrome vertices in \(V\) at time \(t\).
Let \(x\) be the syndrome bitstring.
Since the vertices of \(V\) were picked such that they are independent, the Hamming weight \(|x|\) is a sufficient statistic.
The maximum likelihood estimate of \(q\) is
\begin{align*}
    \hat{q} &= \argmax_{\tilde{q} \in \{p_1,p_2\}} \log \Pr\left(|x| \mid q=\tilde{q}\right) \\
    &= \argmax_{\tilde{q} \in \{p_1,p_2\}} \left[ |x| \log \tilde{q} + (n-|x|) \log (1-\tilde{q}) \right]
\end{align*}
The maximization problem can be conveniently rewritten in terms of the log-likelihood ratio \(\ell(|x|)\).
\begin{align*}
    \ell(|x|) = \left[ |x| \log \frac{p_1}{p_2} + (n-|x|) \log \frac{1-p_1}{1-p_2} \right]
\end{align*}
The value of \(\hat{q}\) is then
\begin{align*}
    \hat{q} = \begin{cases}
        p_1 & \ell(|x|) \ge 0 \\
        p_2 & \ell(|x|) < 0
    \end{cases}
\end{align*}
Let \(D(a||b) = a \log \frac{a}{b} + (1-a) \log \frac{1-a}{1-b}\) be the Kullback-Leibler divergence between two Bernoulli distribution with parameter \(a\) and \(b\), respectively.
Employing a Chernoff bound on \(|x|\), we have an upper bound on the probability of a false positive
\begin{align}
    \Pr(\hat{q} = p_2 \mid q=p_1) &= \Pr(\ell < 0 \mid q=p_1) \\
    &= \Pr\left(\frac{|x|}{n} > \alpha \mid q=p_1\right)\\
    &\le e^{-n D(\alpha || p_1)}
\end{align}
where \(\alpha = \frac{\log \frac{1-p_2}{1-p_1}}{\log{\frac{p_1 (1-p_2)}{p_2(1-p_1)}}} \in (p_1,p_2)\).
Likewise, we can upper bound the probability of a false negative:
\begin{align}
    \Pr(\hat{q} = p_1 \mid q=p_2) &= \Pr(\ell \ge 0 \mid q=p_1) \\
    &= \Pr\left(\frac{|x|}{n} \le \alpha \mid q=p_2\right)\\
    &\le e^{-n D(\alpha || p_2)}
\end{align}
To pick some representative parameters, for \(p = 0.01\), \(p_B=0.1\), \(n=100\), and \(w=10\), the probability of false negatives and false positives is at most \(10^{-4}\).

\subsection{Estimation of threshold values for threshold experiments}\label{sec:estimating_error_burst}
To obtain the phase diagrams in Section~\ref{sec:phase_diagram}, we perform memory experiments ($2\cdot 10^6$ samples) with surface code distances between \(d=13\) and \(d=27\) with $T = 2d$ syndrome extraction cycles and an error burst after the \(d\)-th syndrome extraction cycle.

For the high \(p_B\) portions of the phase diagram, we adopt the same fitting strategy as Wang, et al. \cite{Wang_2003} and perform a fit of the logical error rates as function of \(p_B\) (\(p\) fixed) to a quadratic function in a scaled variable:
\begin{eqnarray}\label{eqn:quadratic}
    P_L(x) = A + Bx + Cx^2
\end{eqnarray}
where $x = \left(p_B - p_B^*\right)d^{1/\nu_0}$ for code distance $d$, constants $A, B, C$ and some critical exponent $\nu_0$. 
We determine the threshold $p_B^*$  by fitting the values to \eqref{eqn:quadratic}.

When the background error rate $p$ is high, we adopt an alternate approach in order to mitigate large finite size effects. 
The procedure is nearly identical to the standard procedure, but \(p_B\) is held fixed and \(p\) is swept.
The fitting ansatz \eqref{eqn:quadratic} is performed with \(p_B\) replaced by \(p\).

\subsection{Low error rate extrapolation}\label{sec:low-error-rate-extrapolation}

\begin{figure}[]
    \centering    \includegraphics[width=0.5\textwidth]{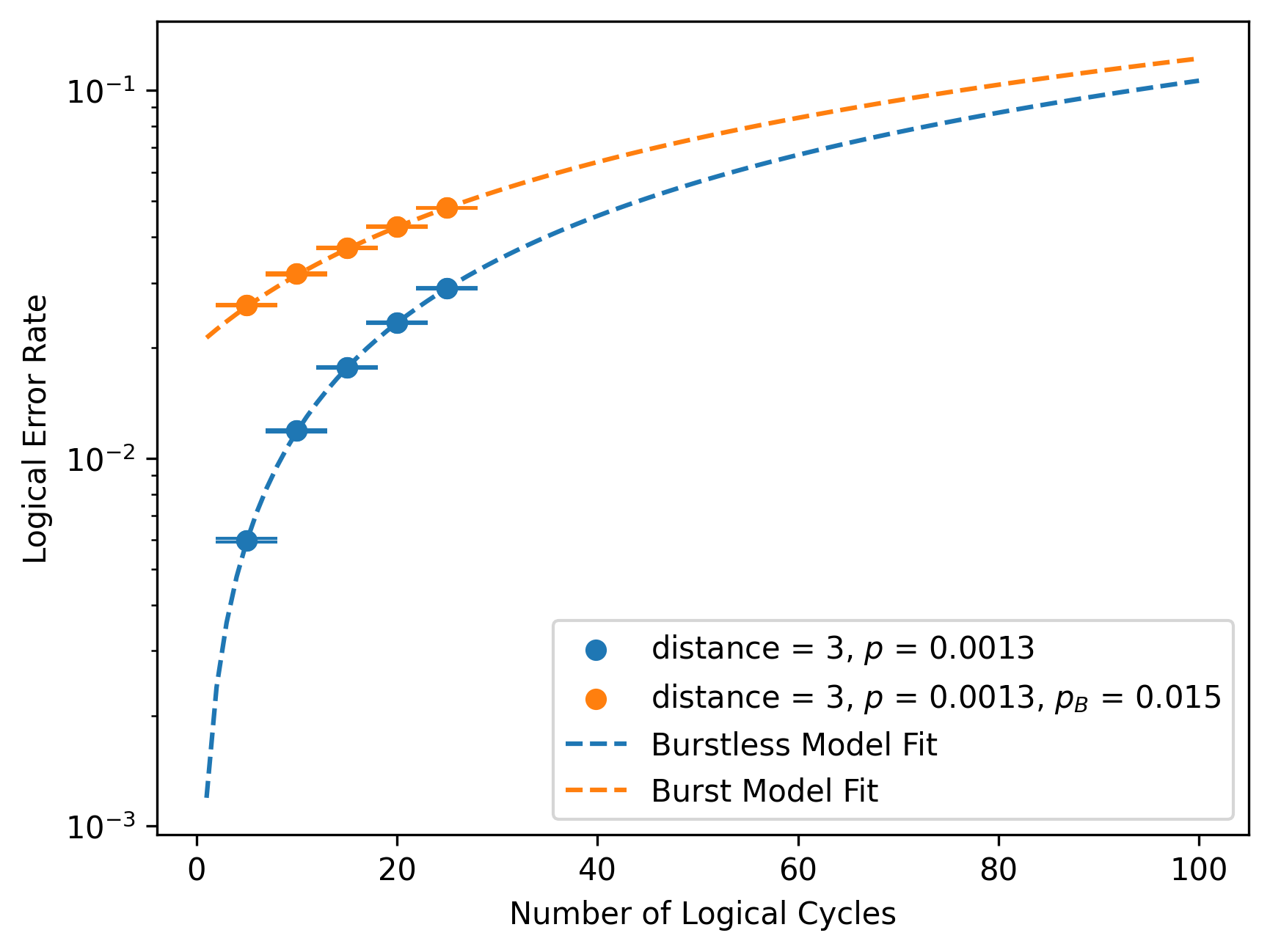}
    \caption{Plot of Overall Logical Error Rate %
    against Number of Logical Cycles %
 for background error rate $p = 0.0013$, burst error rates $p_B=0$ and $p_B = 0.015$, and distance $d = 3$. 
 The dashed lines are obtained with our ansatze stated in \cref{eq:ansatz_p_L} and \cref{eq:ansatz_p_LB} using $q_d$ and $q_{d,B}$ values for $D = 25$ logical cycles.}
    \label{fig:ansatz-fitting}
\end{figure}

\begin{figure}[]
    \centering    
    \includegraphics[width=0.5\textwidth]{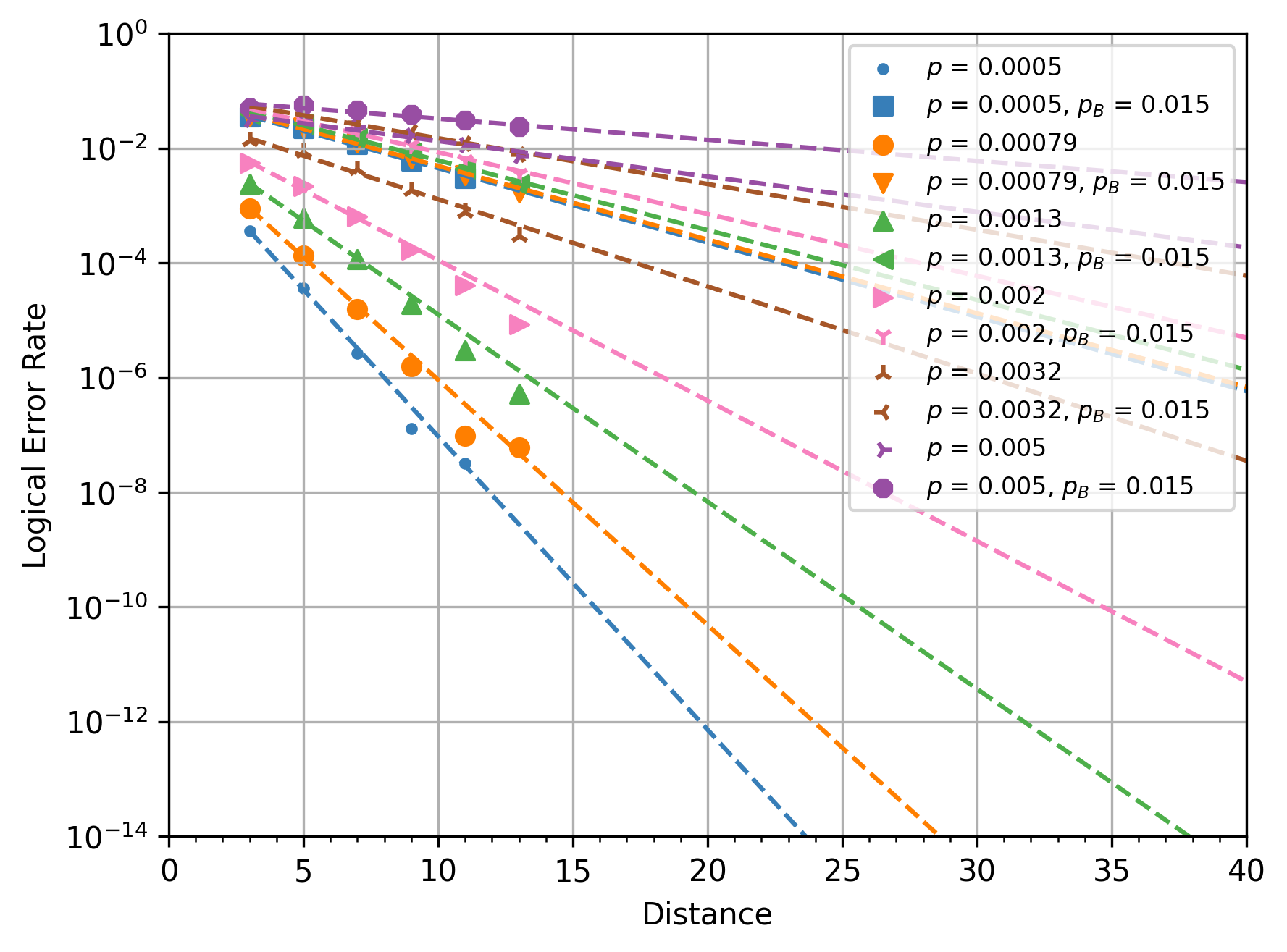}
    \caption{Logical Error Rates for various values of the background error rate $p$, the burst error rate $p_B$, and the code distance $d$.}
    \label{fig:preteraquop}
\end{figure}

To study the subthreshold behavior, we first determine the logical error rates obtained from memory experiments with various values of $p$ and $p_B$ and number of logical cycles $D$ between $D = 5$ and $D = 25$, obtaining the results plotted in \cref{fig:ansatz-fitting}.
We derive the values for $q_{d}$ and $q_{d,B}$ for different values of $d$, $p$, and $p_B$ by using \cref{eq:ansatz_p_L} and \cref{eq:ansatz_p_LB}.
We observe that $q_d$ and $q_{d,B}$ approach constant values as $D %
\to 25$ and set $\tilde{q}_{d} = q_{d}$ and $\tilde{q}_{d,B} = q_{d,B}$ for $D = 25$.
We fit the plotted points corresponding to different $\tilde{q}_d$ and $\tilde{q}_{d,B}$ in \cref{fig:preteraquop} to the expression $\log_{10}\tilde{q}_{d} = c_{\tilde{q}_{d}}+d\cdot m_{\tilde{q}_{d}}$, and similarly for $\log_{10}\tilde{q}_{d,B}$.
The parameters determined by the fits shown in \cref{tab:fit_params} were used for the teraquop footprint estimate.

\newcommand{\ra}[1]{\renewcommand{\arraystretch}{#1}}

\begin{table*}\centering
\ra{1.3}
\begin{tabular}{@{}rrrrr@{}}\toprule

Background error rate $[\%]$ & $c[{\tilde{q}_d}]$ & $m[\tilde{q}_d]$ & $c[\tilde{q}_{d,B}]$  & $m[\tilde{q}_{d,B}]$ \\ \midrule

$0.08$ & $-1.77(3)$  &$-0.421(7)$ & $-0.902(5)$  & $-0.0459(7)$ \\

$0.12$ & $-1.655(13)$  & $-0.343(4)$ &  $-0.872(5)$  & $-0.0459(6)$  \\

$0.18$ & $-1.531(7)$ & $-0.267(2)$  &  $-0.847(5)$  & $-0.0420(6)$ \\

$0.27$ & $-1.419(4)$  & $-0.1856(7)$  & $-0.827(6)$  & $-0.0342(6)$  \\

$0.40$ & $-1.311(2)$  & $ -0.1069(3)$ &  $-0.833(8)$ & $-0.0188(7)$  \\
\bottomrule
\end{tabular}
\caption{Fit parameters for circuit-level depolarizing noise with burst error rate $p_B = 2.5\%$ and five different values of the background error rate $p$. Here $\log_{10} \tilde{q}_d = c[\tilde{q}_d]+ d\cdot m[\tilde{q}_d]$, and $\log_{10} \tilde{q}_{d,B} = c[\tilde{q}_{d,B}]+ d\cdot m[\tilde{q}_{d,B}]$.
All confidence intervals are two standard-deviations. \label{tab:fit_params}}
\end{table*}

\end{document}